\documentclass{amsart}
\usepackage{hyperref}
\usepackage{amsmath, amssymb,graphicx}
\usepackage{epsf}

\DeclareMathOperator{\Cov}{Cov}
\DeclareMathOperator{\tr}{Tr}
\DeclareMathOperator{\diag}{diag}
\DeclareMathOperator{\Aut}{Aut}
\DeclareMathOperator{\ch}{ch}

\newcommand{\beq}{\begin{equation}}
\newcommand{\eeq}{\end{equation}}
\newcommand{\bea}{\begin{eqnarray}}
\newcommand{\eea}{\end{eqnarray}}

\newcommand{\abs}[1]{ \left| #1 \right| }

\newcommand{\CP}{\mathbb{C}\mathrm{P}^1}

\newcommand{\Hermitian}{\mathcal{H}}
\renewcommand{\l}{\mathbf{\lambda}}
\renewcommand{\v}{\mathbf{v}}
\newcommand{\h}{\mathbf{h}}
\newcommand{\p}{\mathbf{p}}

\newcommand{\R}{\mathbf{R}}
\newcommand{\dd}{\mathrm{d}}
\newcommand{\Res}{\mathop{\,\rm Res\,}}

\begin{document}

{\hfill IPHT T09/055}

\title{A matrix model for simple Hurwitz numbers, and topological recursion}

\author[G. Borot]{Ga\"{e}tan Borot}
\author[B. Eynard]{Bertrand Eynard}
\author[M. Mulase]{Motohico Mulase}
\author[B. Safnuk]{Brad Safnuk}

\begin{abstract}
We introduce a new matrix model representation for the generating
function of simple Hurwitz numbers. We calculate the spectral curve
of the model and the associated symplectic invariants developed in
\cite{EOFg}. As an application, we prove the conjecture proposed by
Bouchard and Mari\~{n}o \cite{BM}, relating Hurwitz numbers to the
spectral invariants of the Lambert curve $e^x = ye^{-y}$.
\end{abstract}

\maketitle

\section{Summary}


In \cite{BM}, Bouchard and Mari\~no propose a new conjectural
recursion formula to compute simple Hurwitz numbers, i.e. the
weighted count of coverings of $\CP$ with specified branching data.
Their recursion is based on a new conjectured formalism for the type
B topological string on mirrors of toric Calabi--Yau threefolds,
called ``remodeling the B-model'', or ``\textsc{bkmp} conjecture''
\cite{BKMP}. The Bouchard-Mari\~no conjecture for Hurwitz numbers
appears as a consequence of this general \textsc{bkmp} conjecture
applied to the infinite framing limit of the open string theory of
$\mathbb{C}^3$. In this limit, the amplitudes are known to give
simple Hurwitz numbers.

They propose that the generating function for Hurwitz numbers can be
recovered from the symplectic invariants (also called topological
recursion) developed in \cite{EOFg}, applied to the so called
``Lambert curve'' $y=L(e^x)$ defined by:
$$
  e^x = ye^{-y}.
$$


 In this paper, we make the link between Hurwitz
numbers and the Lambert curve explicit. We introduce a new matrix
model formula for the generating function of simple Hurwitz numbers
\beq
     \label{eq:MatrixModel} Z \propto \int_{\mathcal{H}_N(\mathcal{C})}\mathrm{d}M\exp\left(-\frac{1}{g_s}\tr(V(M) -
     M\R)\right),
\eeq
where $V(x)$ is the potential
$$
V(x) = -\frac{x^2}{2} + g_s(N - \frac{1}{2}) x +x \ln (g_s/t)+ i\pi x
-g_s\ln\bigl(\Gamma(-x/g_s)\bigr).
$$

The parameters $g_s$ and the matrix $\R$ involved in the definition
of $Z$ are such that the weight of a covering of Euler
characteristic $\chi$ is proportional to $g_s^{\chi}$, and has a
polynomial dependance in $v_i = \exp{R_i}$ which encodes the
ramification data above branch points.


A method to compute topological expansion of matrix integrals with
an external field was introduced in \cite{EOFg}. It consists in
finding the spectral curve ${\mathcal S}$ (roughly speaking the
equilibrium density of eigenvalues of the matrix, more precisely the
planar part of the expectation value of the resolvent), then
computing recursively a sequence of algebraic $k$-forms
$\mathcal{W}_k^{(g)}({\mathcal S})$, and some related algebraic quantities
called symplectic invariants ${\mathcal F}_g({\mathcal S}) =
\mathcal{W}_0^{(g)}({\mathcal S})$. Then, one of the main results of
\cite{EOFg} is that
$$
\ln Z = \sum_{g=0}^\infty g_s^{2g-2}\, {\mathcal F}_g({\mathcal S}).
$$
In our case, this implies that the generating function for simple
Hurwitz numbers of genus $g$ is precisely  ${\mathcal F}_g({\mathcal
S})$, where ${\mathcal S}$ is the spectral curve of our matrix
model.

It is rather easy to find the spectral curve of the matrix model
Eqn~\ref{eq:MatrixModel}. The result, after suitable symplectic
transformations, reads
  \beq
\widetilde{\mathcal{S}}(\p,g_s;t) =
\left\{\begin{array}{l}
x(z)= - z  + \ln{(z/t)} + c_0 + \frac{c_1}{z} - \sum_{n=1}^\infty {B_{2n}\, g_s^{2n}\over 2n}\, f_{2n}(z)\\
y(z)= z + g_s\sum_{i=1}^N \frac{1}{(z-z_i)y_i}+\frac{1}{z_i y_i}
\end{array}\right. ,
\eeq
where $z_i$, $y_i$, $c_0$ and $c_1$ are determined by consistency relations, and $z_i$ and $y_i$ are $O(1)$ when $g_s \rightarrow 0$, and $c_0$ and $c_1$ are $O(g_s)$. 
In particular, when we set the coupling
constant $g_s = 0$, we recover the Lambert curve  $y=L(t e^x)$.

In \cite{BM}, Bouchard and Mari\~no define another set of generating
functions, denoted $H^{(g)}(x_1,\ldots,x_k)$, encoding genus $g$ simple
Hurwitz numbers, and which are derivatives of $\ln Z$, evaluated at $g_s=0$.
The statement of their conjecture is:
$$
   \frac{{\mathcal W}_k^{(g)}(z_1,\ldots,z_k)}{\mathrm{d}x(z_1)\cdots\mathrm{d}x(z_k)}
  =  H^{(g)}(x(z_1),\ldots,x(z_k)),
$$
where ${\mathcal W}_k^{(g)}$ are the k-forms of \cite{EOFg} computed for the Lambert curve.

We prove their conjecture by using the general properties of the invariants of \cite{EOFg}, in particular the fact that derivatives of the ${\mathcal F}_g$'s with respect to almost any parameter, can be expressed in terms of the ${\mathcal W}_k^{(g)}$'s. Then it suffices to set $g_s=0$, and this gives the Bouchard-Mari\~no conjecture.

\subsection*{Organization of the paper} In Section~\ref{sec:Construction}, we recall the definitions and derive
a matrix model formula for the generating function of simple Hurwitz
numbers. We recall the construction of the symplectic invariants and
topological recursion of \cite{EOFg} in
Section~\ref{sec:Matrixmodels}. In Section~\ref{sec:SpectralCurve},
we derive the spectral curve of our matrix model (the proof is
presented in Appendix \ref{app:SpectralDerivation}) and prove the
Bouchard-Mari\~no conjecture following a method very close to
\cite{MR2439683}.
In section~\ref{sec:ELSV}, we briefly study the link with the Kontsevich integral.
Section~\ref{sec:conclusion} addresses
generalizations of our method and open questions.


\section{Construction of the matrix model}
\label{sec:Construction}

Let $\Cov^*_n(C_1, \ldots, C_k)$ denote the weighted number of $n$-fold  coverings (possibly disconnected) of $\CP$ ramified over $k$ fixed points of $\CP$ with monodromies in the conjugacy classes $C_1, \ldots, C_k$. The weight is one over the order of the automorphism group of the covering.
Similarly, let $\Cov_n(C_1, \ldots, C_k)$ denote the weighted number of $n$-fold {\bf connected} coverings.

By a result of Burnside (see, eg \cite{MR1783622}), we have 
\beq
\label{eq:Burnside}\Cov^*_n(C_1, \ldots, C_k) =
\sum_{\abs{\mathbf{\lambda}} = n} \left(
\frac{\dim\mathbf{\lambda}}{n!} \right)^2 \prod_{i=1}^k
f_{\mathbf{\lambda}}(C_i), \eeq
where the sum ranges over all
partitions $\mathbf{\lambda} = (\lambda_1 \geq \lambda_2 \geq \cdots
\geq \lambda_n \geq 0)$ of $\abs{\mathbf{\lambda}} = \sum{\lambda_i}
= n$ boxes, $\dim\mathbf{\lambda}$ is the dimension of the
irreducible representation indexed by $\mathbf{\lambda}$ (with
corresponding character $\chi_{\mathbf{\lambda}}$), and
$$
  f_{\mathbf{\lambda}}(C_i) = \frac{\abs{C_i}}{\dim\mathbf{\lambda}}
  \chi_{\mathbf{\lambda}}(C_i).
$$

\subsection{Simple Hurwitz numbers}

\begin{figure}[h]
 \centering
  \includegraphics[width=14cm]{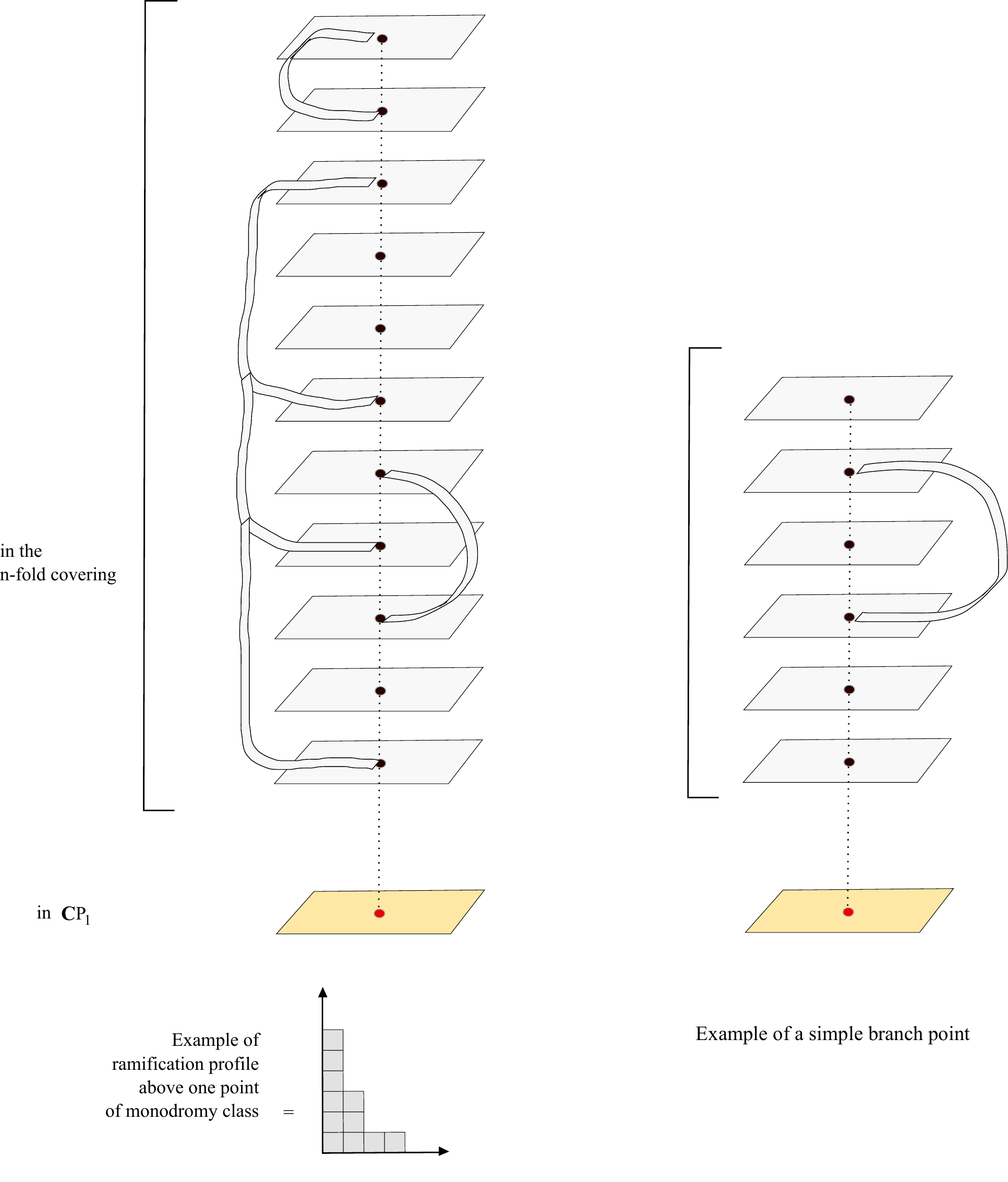}
 \caption{\footnotesize{Branched covering, with one branch point of monodromy class $\mu$, and simple branch points (for which the monodromy is a transposition).}}
 \label{fimonodromies}
\end{figure}

We are interested in counting $n$-fold coverings of genus $g$ with 1
branch point of arbitrary profile $\mu$, and only transpositions
above other points (called simple branch points), c.f.
Figure~\ref{fimonodromies}. We denote $C_{(2)}$ the conjugacy class
of a transposition. For $b$ simple branch points and one branch
point of profile $\mu$, the Euler characteristic of the $n-$fold
covering reads from the Riemann-Hurwitz formula:
$$
\chi = |\mathbf{\mu}| + \ell(\mathbf{\mu})- b.
$$
For connected coverings, we have $\chi=2-2g$, where $g$ is the
genus, and we also define the simple Hurwitz numbers:
$$
H_{g,\mathbf{\mu}} =  \Cov_{n}(C_{\mathbf{\mu}}, \overbrace{C_{(2)}, \ldots, C_{(2)}}^{b} ) ,
$$
where  $b = 2g - 2 + |\mathbf{\mu}| + \ell(\mathbf{\mu})$.

\subsection{Generating function for simple Hurwitz numbers}
\label{sec:Topoexp}

With the notations $p_{\mathbf{\mu}} = \prod_i p_{\mu_i}$, and $\mathbf{p} = (p_1,p_2, \ldots)$, we shall study the generating function
\begin{equation}
\label{Z}
  Z\left(\mathbf{p},g_s;t\right) = \sum_{n=0}^\infty t^n\,\,\,  \sum_{|\mathbf{\mu}|=n}
\,\,\sum_{b=0}^\infty
\frac{g_s^{b-|\mathbf{\mu}|-\ell(\mathbf{\mu})}}{b!}
\,\,p_{\mathbf{\mu}} \Cov^*_{n}(C_{\mathbf{\mu}},
\overbrace{C_{(2)}, \ldots, C_{(2)}}^b ) ,
\end{equation}
where $\Cov^*_{n}(C_{\mathbf{\mu}},C_{(2)}, \ldots, C_{(2)} )$ is
the number of (not necessarily connected) branched coverings of
Euler characteristic $\chi = |\mu|+\ell(\mu)-b$.

In the language of string theory, $-g_s$ is the string coupling constant\footnote{In the physics literature, it is customary to choose $-g_s$ instead of $g_s$ as formal parameter.}. Let us emphasize that $Z(\p,g_s;t)$ is defined as a formal power series in $t$ and $g_s$, i.e. it is merely a notation to collect all the coefficients. Each coefficient (for $b$ and $n$ fixed) is a finite sum, which is a polynomial function of $p_1,\dots, p_n$. Notice also that the parameter $t$ is redundant because we can change $p_j\to \rho^j p_j$ and $t\to t/\rho$ without changing the sum, i.e.
$$
Z(\{p_1,p_2,p_3,\dots,\},g_s;t)=Z(\{\rho p_1, \rho^2 p_2, \rho^3 p_3,\dots,\},g_s;t/\rho).
$$
In \cite{BM}, $t$ is chosen as $t=1$, but we find more convenient to
keep $t\neq 1$ for the moment, in order to have only two formal
parameters $g_s$ and $t$, instead of an infinite number of them
$g_s$ and $p_1,p_2,\dots$, which would be the case if $t$ were set
to 1. The generating function of {\bf connected} coverings is $F =
\ln Z$ (in the sense of formal power series of $t$ and $g_s$):
$$
F(\mathbf{p}, g_s;t) = \ln Z =  \sum_{b,n} \frac{t^n}{b!} \sum_{\abs{\mathbf{\mu}} = n} g_s^{2g-2}\,\,p_{\mathbf{\mu}} H_{g, \mu}
,$$
where $b = 2g - 2 + |\mathbf{\mu}| + \ell(\mathbf{\mu})$.

Therefore, we have a so-called \emph{topological expansion} (equality of
formal series):
$$
F(\mathbf{p},g_s;t) = \sum_{g=0}^\infty g_s^{2g-2}\,\, F_g(\mathbf{p};t),
$$
where $F_g$ counts the number of connected coverings of genus $g$:
$$
F_g(\mathbf{p};t) =  \sum_{n}\, t^n  \sum_{|\mathbf{\mu}|=n} \frac{p_\mathbf{\mu}}{(2g - 2 + n + \ell(\mathbf{\mu}))!} \,H_{g,\mathbf{\mu}}.
$$
Our goal in this article is to provide a recursive algorithm to compute the $F_g$'s, and more precisely, prove that the $F_g$'s are the symplectic invariants introduced in \cite{EOFg} for a spectral curve $\widetilde{\mathcal{S}}(\p,g_s;t)$ which we shall describe in Section~\ref{sec:SpectralCurve}. As a consequence, we shall prove the conjecture of Bouchard and Mari\~no \cite{BM}.

\subsection{Partitions}

To make our notations clear, we recall some representations of
partitions or Young tableaux. The set of all partitions $\lambda$ of
length $\leq N$
 is in bijection with other interesting sets of
objects:
\begin{itemize}
\item[$\bullet$] The decreasing finite series of $N$ integers : $\lambda_1 \geq \ldots \geq \lambda_{\ell(\l)} \geq \lambda_{\ell(\l) + 1} = \ldots = \lambda_N\geq  0$.
The length $\ell(\l)$ of the partition is the number of
non-vanishing $\l_i$'s.
$|\l| = \sum_{i = 1}^{\ell(\l)} \l_i = n$ is the number of boxes of the partition. \\
\item[$\bullet$] The strictly decreasing finite series of positive integers. They mark the positions (up to a translation) on the horizontal axis of the increasing jumps when the Young tableau $(\lambda_i,i)$ is tilted anticlockwise by $\frac{\pi}{4}$ (c.f. Figure~\ref{figYoung}). The correspondence is given by
$$
h_i = \lambda_i - i + N \qquad(i \in
\{1,\ldots,N \}).
$$
We have $h_1>h_2>\dots>h_N\geq 0$.

\begin{figure}[h]
 \centering
  \includegraphics[width=10cm]{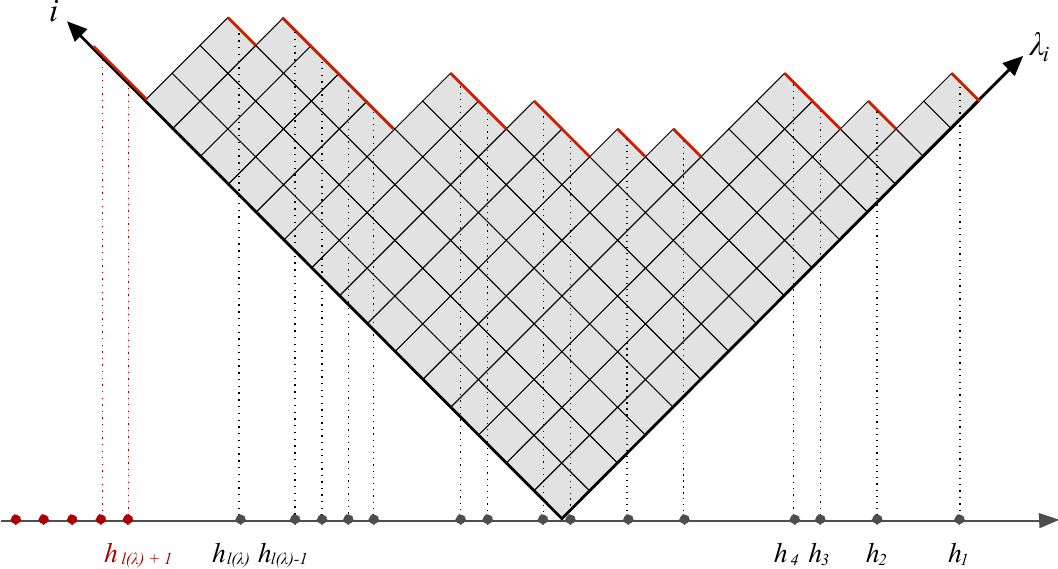}
 \caption{\footnotesize{Rotate the partition by $\pi/4$. The $h_i$'s mark the positions (up to a translation) on the horizontal axis of the increasing jumps in the Young tableau.}}
 \label{figYoung}
\end{figure}

\item[$\bullet$] The conjugacy classes of $\mathfrak{S}_n$. The class
$C_{\l}$ associated to $\l$ is the one
with $m_r = |\{i > 0 \quad \lambda_i = r\}|$ cycles of length $r$, and
its cardinal is
$$
|C_{\mathbf{\lambda}}| = \frac{|\lambda|!}{\prod_{r \geq 1} m_r!\cdot r^{m_r}}.
$$ 
\item[$\bullet$] The equivalence classes of irreducible representations of $\mathfrak{S}_{|\mathbf{\lambda}|}$.
\item[$\bullet$] The equivalence classes of irreducible representations of $\mathrm{GL}_{\ell(\mathbf{\lambda})}(\mathbb{C})$, or of $\mathrm{U}(\ell(\mathbf{\lambda}))$.
\end{itemize}

\subsection{Schur polynomials}

Recall the definition of Schur polynomials $s_{\mathbf{\lambda}}$ \cite{MR1354144}: they coincide with the characters of the representation of $\mathrm{U}(k)$ indexed by $\mathbf{\lambda}$ such that $k = \ell(\mathbf{\lambda})$. As a matter of fact, if $\mathbf{v} = (v_1,\dots,v_k)$ is an $k$-tuple of complex variables:
\beq
\label{Schur}
s_{\mathbf{\lambda}}(\mathbf{v}) = \frac{\det\left(v_i^{\mathbf{\lambda}_j - j + N}\right)}{\Delta(\mathbf{v})},
\eeq
where $\Delta(\mathbf{v})=\prod_{1\leq j<i\leq N} (v_i-v_j)$ is the Vandermonde determinant. This formula can be extended to a definition of $s_{\lambda}$ with $N \geq k$ variables, $\mathbf{v} = (v_1,\ldots,v_N)$, provided that we take $\lambda_j = 0$ whenever $j > k$.

The Frobenius formula gives the expansion of a Schur polynomial in
terms of the power-sum functions \mbox{$\widetilde{p}_m =
\sum_{i=1}^{N} v_i^m$}, \beq
 \label{Frob1} s_{\mathbf{\lambda}} (\mathbf{v}) = \frac{1}{n!}
\sum_{\abs{\mathbf{\mu}} = n}\abs{C_{\mathbf{\mu}}}\chi_{\mathbf{\lambda}}(C_{\mathbf{\mu}})\,\,\widetilde{p}_{\mu},
\eeq which stresses the link between $\mathrm{U}(k)$ characters and
$\mathfrak{S}_{n}$ characters. It is still valid with $N \geq k$ variables instead of $\ell(\mathbf{\lambda}) = k$ variables.

From Eqn~\ref{Schur}, one can obtain the classical result for the
dimension of the representation indexed by $\mathbf{\lambda}$:
$$
s_{\mathbf{\lambda}}(1,\ldots,1) = \mathrm{dim}\;\mathbf{\lambda} =
\frac{\Delta(\mathbf{h})}{\prod_{i = 1}^{N} h_i !},
$$
while the Frobenius formula leads to the expression
$$
f_{\mathbf{\lambda}}(C_{2}) = \frac{1}{2}\sum_i h_i^2 - (N -
\frac{1}{2})\sum_i h_i + \frac{N}{3}(N^2 - \frac{3}{2}N + 2).
$$

\subsection{$Z$ as a sum on partitions}

After Eqn~\ref{Frob1}, if we consider the variables $p_m$'s to be
power-sum functions of some $N$-upple parameter $\v$, we have \beq
\left\{\:\begin{array}{rcl} Z(\mathbf{p}, g_s;t) & = & \sum_{b,n}  \frac{t^{n}\, g_s^{b-n}}{b!} \sum_{\ell(\l)\leq N, |\l|=n} \frac{\dim\:\l}{n!}\,\, {s_{\l}(\mathbf{v})} \left(f_{\l}(C_{(2)})\right)^b \\
p_m & = & g_s\sum_{i = 1}^N v_i^m. \end{array}\right.
\eeq
Hence,
$$
   Z(\mathbf{p}, g_s;t) =  \sum_{\ell(\l)\leq N} (t/g_s)^{|\l|}\,\, \frac{\dim\:\l}{|\lambda|!}\,\, {s_{\l}(\v)} \,\,\,e^{g_s  f_{\mathbf{\lambda}}(C_{(2)})}.
$$
Alternatively, given that $s_{\l}$ is homogeneous of degree $|\l|$, we have
$$
   Z(\mathbf{p}, g_s;t) =  \sum_{\ell(\l)\leq N}  \frac{\dim\:\l}{|\l|!}\,\, {s_{\l}(t\v/g_s)} \,\,\,e^{g_s  f_{\l}(C_{(2)})}.
$$

Again, we emphasize that the above expression for $Z$ ought to be considered as a formal
power series in $t$ and $g_s$. For a given $b$ and $n$, the
coefficient of $t^n g_s^{\chi}$ is a polynomial in the $p_m$'s,
which involves only $p_1,\dots, p_n$. Therefore, it is always
possible to find $N$ (independent of $b$) and $v_1,\ldots,v_N$ such
that $\forall i \in \{1,\ldots,N\}$ we have $p_m = g_s\sum v_i^m$.
The values of the $p_m$'s for $m > n$ do not matter.

\subsection{Matrix integral representation of $Z$}

To write $Z$ as a matrix integral we express $s_{\l}$ in terms of
the Itzykson-Zuber integral \cite{MR562985}:
\bea
    \mathrm{I}(X,Y) & = &  \int_{\mathrm{U}(N)} \mathrm{d}U\,e^{ \tr(XUYU^{\dagger})} \nonumber \\
      & = & \frac{\det(e^{ x_i y_j }) }{\Delta(X) \Delta(Y)}.
      \nonumber
\eea
Here $dU$ is the Haar measure on $U(N)$, normalized such that the second line holds without any constant prefactor.

Therefore, if we let $\mathbf{R} = \diag(\ln v_1,\ldots,\ln v_N)$
and $\mathbf{h}_{\l} = \diag(h_1, \ldots, h_N)$, we have from
Eqn~\ref{Schur} :
$$
  s_{\l}(\v) = \Delta(\h_{\l})\frac{\Delta(\R)}{\Delta(\v)} I(\h_{\l}, \R).
$$

Then, the partition function looks like
\bea
  Z(\p, g_s;t) & = & \frac{\Delta(\R)}{\Delta(\v)}\sum_{\lambda} I(\h_{\l}, \R) \,\,\frac{\bigl(\Delta(\h_{\l})\bigr)^2} {\prod_{i = 1}^N h_i!} \prod_{i = 1}^N e^{ g_s  A_2(h_i)}\,(g_s/t)^{-A_1(h_i)}  \nonumber\\
  & =  & \frac{\Delta(\R)}{\Delta(\v)} \sum_{h_1 > \cdots > h_N \geq 0} I(\h, \mathbf{R}) \,{\bigl(\Delta(\h)\bigr)^2}\prod_{i = 1}^N \frac{e^{ g_s  A_2(h_i)}(g_s/t)^{-A_1(h_i)}}{\Gamma(h_i + 1)} \nonumber  \\
  & = & \frac{1}{N!}\frac{\Delta(\R)}{\Delta(\v)} \sum_{h_1, \ldots, h_N \geq 0} I(\h,\R)\,\bigl(\Delta(\h)\bigr)^2\prod_{i = 1}^N \frac{e^{ g_s  A_2(h_i)} (g_s/t)^{-A_1(h_i)}}{\Gamma(h_i + 1)}, \nonumber
\eea
\noindent
 where $|\l| = \sum_i A_1(h_i)$ and $f_{\l}(C_{(2)}) = \sum_i
A_2(h_i)$. The factorization of the weight with respect to the $h_i$'s and the presence
of a squared Vandermonde is the key for the representation of $Z$ as a
hermitian matrix model.

\begin{figure}
\includegraphics[width=\textwidth]{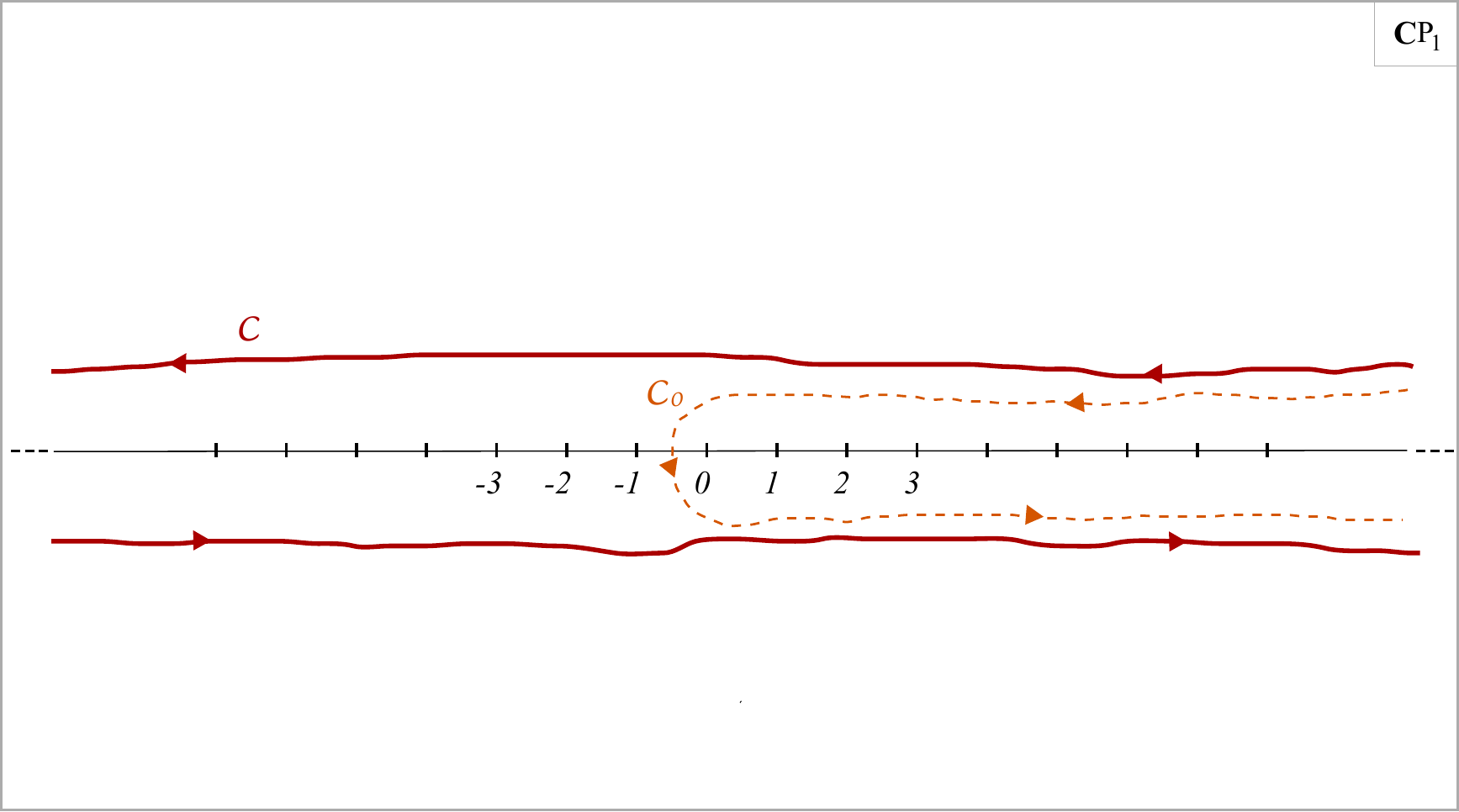}
\caption{Contours enclosing the non-negative integers and the entire real line.}
\label{fig:Contour}
\end{figure}

 As it was done in \cite{MR2439683}, we
represent the $N$ sums as integrals over a contour, namely the
contour $\mathcal{C}_0$ enclosing the non-negative integers, as
pictured in Figure~\ref{fig:Contour}.
We make use of a function which has simple poles with residue 1 at all integers:
$$
  f(\xi) = \frac{\pi e^{-i\pi \xi}}{\sin(\pi \xi)} = - \Gamma(\xi + 1)\Gamma(-\xi)e^{-i\pi\xi}.
$$
Thus, we have:
\bea
Z(\p, g_s;t) 
&=&   \frac{1}{N!}\frac{\Delta(\R)}{\Delta(\v)}
\oint_{\mathcal{C}_0^N} d h_1\cdots d h_N \bigl(\Delta(\h) \bigr)^2
I(\h, \R) \nonumber \cr 
&& \qquad \prod_{i=1}^N \frac{f(h_i)e^{ g_s A_2(h_i)}\,
(g_s/t)^{-A_1(h_i)}}{\Gamma(h_i + 1)}. \nonumber 
\eea
Actually, the ratio $f(\xi)/\Gamma(\xi + 1)$ has only poles at non-
negative integers, it has no pole at negative integers. So, we can
replace $C_0$ by the contour $\mathcal{C}$ from
Figure~\ref{fig:Contour} which encloses all integers, and which we
choose invariant under translation on the real axis. Therefore, we
arrive at
\bea
Z(\p, g_s;t) 
&=&   \frac{1}{N!}\frac{\Delta(\R)}{\Delta(\v)}
\oint_{\mathcal{C}^N} d h_1\cdots d h_N \bigl(\Delta(\h) \bigr)^2
I(\h, \R) \nonumber \cr 
&& \qquad \prod_{i=1}^N \frac{f(h_i)e^{ g_s A_2(h_i)}\,
(g_s/t)^{-A_1(h_i)}}{\Gamma(h_i + 1)}. \nonumber 
\eea
The set $\Hermitian_N({\mathcal C})$ of normal matrices with eigenvalues on ${\mathcal C}$ is the set of matrices $M$ which can be diagonalized by conjugation with a unitary matrix, and whose eigenvalues belong to ${\mathcal C}$:
$$
M = U^\dagger \, X\, U , \ U U^\dagger = {\rm Id} , \ X={\rm diag}(x_1,\dots,x_N) , \  x_i\in {\mathcal C}.
$$
$\Hermitian_N({\mathcal C})$ is endowed with the measure
$$
\dd M= \Delta(X)^2\,\, \dd X\, \dd U,
$$
where $\dd U$ is the (up to normalization) Haar measure on $U(N)$
and $\dd X$ is the product of Lebesgue curvilinear measures along
${\mathcal C}$.

From the above discussion, we can express our
generating function as the matrix integral
$$
Z(\p, g_s;t)= \frac{1}{N!}\frac{\Delta(\R)}{\Delta(\v)}
\int_{\Hermitian_N(\mathcal{C})} \mathrm{d}M e^{-\tr \hat V(M) +
\tr(M\R)},
$$
where
$$
\hat V(\xi) = - g_s A_2(\xi)+\ln (g_s/t)\,A_1(\xi) + i\pi \xi -
\ln{\bigl(\Gamma(-\xi) \bigr)}.
$$
Since we are interested in the expansion as a power series in $g_s$, we prefer to rescale $\xi=x/g_s$, i.e. $M\to M/g_s$, and rewrite
\beq
\boxed{\label{eq:ZZ}Z(\p, g_s;t)= \frac{g_s^{-N^2}}{N!}\frac{\Delta(\R)}{\Delta(\v)}
\int_{\Hermitian_N (\mathcal{C})} \mathrm{d}M e^{-\frac{1}{g_s}\tr
\bigl(V(M)-M\R \bigr)},}
\eeq
\noindent where now the potential reads
\begin{align*}
V(x) & =  - g_s^2A_2(x/g_s) + g_s\ln (g_s/t)A_1(x/g_s)+ i\pi x - g_s\ln{\bigl(\Gamma(-x/g_s )\bigr)} \\
& =  - \frac{x^2}{2} +  g_s(N-\frac{1}{2})x + (\ln(g_s/t) + i\pi)x - g_s\ln{\bigl(\Gamma(-x/g_s)\bigr)} + C_t,  \\
\intertext{with}
 C_t & =  - \frac{1}{3}g_s^2(N^2-\frac{3}{2}N + 2) +
\frac{1}{2}g_s(N-1)\ln (g_s/t)
 \end{align*}
 To write its
derivative, let $\psi = \Gamma'/\Gamma$, then
$$
V'(x) = -  x + g_s(N-\frac{1}{2})  + \ln (g_s/t) +  i\pi + \psi(-x/g_s).
$$
We have the well known formula that:
$$
\psi(\xi) = \gamma+{1\over \xi} + \sum_{k=1}^\infty \left({1\over \xi+k}-{1\over k}\right)
$$
where $\gamma$ is the Euler-Mascheroni constant.
In other words, $\psi(\xi)$ has simple poles at all negative integers, and an essential singularity (log singularity) at $\xi=\infty$.
This gives:
\beq\label{eq:Vsumpoles}
V'(x) = -x + g_s(N-\frac{1}{2})  + \ln (g_s/t) +  i\pi + \gamma -{1\over x} - g_s  \sum_{k=1}^\infty \left({1\over x-k g_s}+{1\over k g_s}\right).
\eeq
$V'(x)$ has simple poles at $x=X_j = j g_s$ for all $j\in \mathbb N$, and an essential singularity at $x=\infty$.

Using  Stirling's formula for the large $\xi$ asymptotic expansion of $\psi$
$$
\psi(\xi) \mathop{=}_{\xi \rightarrow \infty} \ln\xi -
\frac{1}{2\xi} - \sum_{l \geq
1}\frac{B_{2l}}{2l}\frac{1}{\xi^{2l}},
$$
we have order by order in the small $g_s$ expansion
\beq\label{eq:VStirling}
\boxed{V'(x) = -  x + \ln {(x/t)} +  g_s(N-\frac{1}{2})  +
\frac{g_s}{2x} - \sum_{l=1}^\infty
\frac{B_{2l}}{2l}\frac{g_s^{2l}}{x^{2l}},}
\eeq
where the $B_l$'s are the Bernoulli number. We recall their generating function
$$ \frac{\xi}{e^{\xi} - 1} = 1 + \sum_{l = 1}^{\infty}\frac{B_l}{l!}\xi^l .$$
The first few are $B_2 = 1/6,\ B_{4}
= -1/30,\ldots$ while $B_{2l + 1} = 0$ for $l \geq 1$.


\section{Generalities on matrix models}
\label{sec:Matrixmodels} \vspace{0.2cm}

Matrix integrals of the form
$$
Z=\int_{\Hermitian_N (\mathcal{C})} \mathrm{d}M e^{-\frac{1}{g_s}
\tr \bigl[ V(M)-M\R \bigr]}
$$
are called "1-matrix model in an external field".
They can be computed for any potential $V$, any contour ${\mathcal C}$
and any external matrix $\R$. Here, our task is even simpler because
we regard this integral as a formal integral, i.e. a formal power
series in powers of $t$ and $g_s$, and therefore all our
computations are to be performed order by order in powers of $t$ and
$g_s$.

In our case $V$ depends on $g_s$, but for the moment, let us assume that $V$ is an arbitrary potential, and in particular we assume that there is no relationship between the coefficients of $V$ and $g_s$.

Typically, in our case (see Eqn \ref{eq:Vsumpoles}), we choose $V$ of the form:
\beq\label{eq:typicalV}
V'(x) = -x + C + \sum_{j=1}^n {u_j\over x-x_j},
\eeq
where for the moment we assume that there is no relationship between the coefficients of $u_j, x_j$ of $V$ and $g_s$.

\subsection{Topological expansion}

For some choices of $V$, $\R$ and ${\mathcal C}$, it may or may not
happen that the convergent integral $Z$ has a power series expansion
in $g_s$
$$
\ln{Z} = \sum_{g=0}^\infty g_s^{2g-2} F_g.
$$
This happens only if ${\mathcal C}$ is a ``steepest descent path''
for the potential $V$ and $\R$. In general, it is rather difficult
to compute the steepest descent paths of a given arbitrary
potential.

Fortunately, when $Z$ is defined as a formal integral we don't need
to find the steepest descent paths, and very often, formal series do have a topological
expansion almost by definition, order by order in the formal parameters, which is the
case here as we argued in Paragraph~\ref{sec:Topoexp}. In other words,
$$
\ln{Z(\p, g_s;t)} = \sum_{g=0}^\infty g_s^{2g-2} F_g(\p;t)
$$
holds order by order in powers of $g_s$ and  $t$.

\subsection{Loop equations and spectral curve}
\label{sec:SCcar}

We introduce the resolvent $W_1$ and auxiliary quantities $P_1$ and $P_{i,j}$:
\bea
\label{eq:W1}
W_1(x) & = & \Bigl< \tr {1\over x-M} \Big> = \sum_{g = 0}^{\infty} g_s^{2g - 1}\,W_1^{(g)}(x) \nonumber \\
P_1(x,y) & = & \Bigl< \tr {V'(x)-V'(M)\over x-M}\,\,{1\over y-\R}
\Big> = \sum_{g=0}^{\infty} g_s^{2g - 1}\,P_1^{(g)}(x,y), \nonumber \\
P_{i,j} & = & \Bigl< ((x_j-M)^{-1})_{i,i}\,\Big> = \sum_{g=0}^{\infty} g_s^{2g - 1}\,P_{i,j}^{(g)}, \nonumber
\eea
which we assume to have topological expansions in powers of $g_s^{2g-1}$.
In our case, their precise definition, order by order in $t$ as a power
series of $g_s$, is given in Appendix~\ref{app:SpectralDerivation}.

Loop equations, also called Schwinger-Dyson equations, is a general technique, which merely reflects the fact that an integral is invariant by change of variable.
It is a standard matrix model exercise, (see \cite{EOFg} for the 1-matrix model in an external fiels), to prove that the invariance
of $Z$ under the infinitesimal change of variable
$$ M \mapsto M + \frac{\epsilon}{x - M}\frac{1}{y - \R} + O(\epsilon^2)$$
implies that the following \emph{loop equation} is satisfied:
\beq \label{Loopeq}
\left\{\begin{array}{l} W_1^{(0)}(x) = P_1^{(0)}(x,Y(x))
\\ Y(x) = V'(x) - W_1^{(0)}(x). \end{array}\right.  \eeq
Notice that $P_1^{(0)}(x,y)$ is a rational function of $y$, of degree $N$.
If $V'$ were a rational function of $x$ (finite $n$ in Eqn \ref{eq:typicalV}),
then $P_1^{(0)}(x,y)$ would be a
rational function of $x$ of degree $n$:
$$
P_1^{(0)}(x,y) = -g_s \tr {1\over y-\mathbf R} + \sum_{j=1}^n\sum_{i=1}^N {u_j\over x-x_j}\,{1\over y-R_i}\,\, P_{i,j}^{(0)}.
$$
Thus the loop equation would
be an algebraic equation, i.e. $Y(x)$ would be an algebraic function
of $x$. Determining the rational function $P_1^{(0)}(x,y)$ (i.e. determining all the coefficients $P_{i,j}$) is
possible but very tedious, and in fact, it is better to characterize
an algebraic function $Y(x)$ by its singularities and its periods. In
general, one would find that $W_1^{(0)}(x)=V'(x)-Y(x)$ has no
singularity at the singularities of $V'$, and the inverse function
$x(Y)$ has poles of residue $g_s$ at the eigenvalues of $\R$. The
genus of the algebraic function $Y(x)$ and the periods $\oint Ydx$
are related to the integration path ${\mathcal C}$.

In general, the relationship between the periods and ${\mathcal C}$
is quite complicated, but, for many applications to combinatorics,
we are considering only formal matrix integrals, i.e formal
perturbation with parameter $t$ of a gaussian integral. In that
case, the spectral curve $Y(x)$ is always a genus $0$ curve (for the Hurwitz matrix integral, we prove it in Appendix \ref{app:SpectralDerivation}). This
means that there exists a parametrization of $Y(x)$ with a complex
variable $z$:
$$
Y(x) \leftrightarrow \left\{\begin{array}{l} x=x(z) \\ Y=y(z),
\end{array}\right.
$$
where $x$ and $y$ are two analytical functions of $z$. The functions $x(z)$ and $y(z)$ are monovalued functions of $z$, but $Y(x)=y(z)$ is multivalued, because there might exist several $z$ such that $x(z)=x$. One of the determinations of $Y(x)$ is called the physical sheet.

The functions $x$ and $y$ are fully determined by their
singularities. More precisely:
\begin{itemize}
\item[$\bullet$] $W_1^{(0)}(x)=V'(x)-Y(x)$ is analytical in the physical sheet; it can have no singularity except at branchpoints.

\item[$\bullet$] From the definition of $W_1^{(0)}$, we have that at $x = \infty$ in the physical sheet
$$W_1^{(0)}(x)\sim \frac{Ng_s}{x(z)} + o(\frac{1}{x(z)}).$$

\item[$\bullet$] As a consequence of Eqn~\ref{Loopeq}, $x(z)$ must have simple poles when $z \to z_i$ such that $y(z_i)=R_i$, i.e.
$$
x(z) \sim {g_s\over (z-z_i)\,\,y'(z_i)}.
$$
\end{itemize}

In our case, for a potential of type \ref{eq:typicalV}, this implies:
\beq\label{genspcurve}
\left\{\begin{array}{l}
x(z)= z + C - g_s\sum_{i=1}^N \frac{1}{(z-z_i)\,y'(z_i)} \\
y(z)= - z +\sum_{j=1}^n \frac{u_j}{(z-\hat z_j)\,x'(\hat z_j)}
\end{array}\right.
\eeq
where $x(\hat z_j)=x_j$, $y(z_i)=R_i$.

The above characterization of the spectral curve $Y(x)$ is valid even if
the potential $V'$ is not rational, for instance if $n\to \infty$.
From now on, the function $Y(x)$, or more precisely the pair of analytical functions
$$
{\mathcal S}= z\mapsto(x(z),y(z))
$$
is called the ``spectral curve'' of our matrix model.

\subsection{Topological recursion}
\label{sec:Toporec}

We recall in this section the construction of the topological recursion from \cite{EOFg}. For our purposes, we only deal with spectral curves of genus
$0$. Hence, we adapt the definitions of \cite{EOFg} to this case.

A \emph{spectral curve} $\mathcal{S}$ is a pair $(x,y)$ of
analytic functions on $\mathbb{C}\mathrm{P}_1$.
Let $a_i$ be the zeroes of $\dd x$, and assume they are simple.
Then, locally at $a_i$, $y \propto \sqrt{(x - x(a_i))}$. Let us
denote $\overline{z}\neq z$, the unique point corresponding to the other branch of the squareroot, such that $x(z) =x(\overline{z})$. $\overline{z}$ is defined locally near the $a_i$'s.

A tower of $k$-forms $\mathcal{W}_k^{(g)}(z_1,\ldots,z_k)$ is constructed as
follows.

\begin{itemize}

\item[$\bullet$] $\mathcal{W}_1^{(0)} = -y\dd x$. 
\medskip

\item[$\bullet$] $\mathcal{W}_2^{(0)}$ is defined as the Bergman\footnote{After Stefan Bergman (1895-1977), mathematician of Polish origin.}
kernel:
$$
\mathcal{W}_2^{(0)}(z_1,z_2) = B(z_1,z_2) = \frac{\dd z_1 \dd z_2}{(z_1 -
z_2)^2}.
$$ 
\medskip

\item[$\bullet$] We define a recursion kernel
\beq
 K(z',z) \frac{-\frac{1}{2}\int_{\overline{z}}^{z}B(z',\cdot)}{(y(z) -
y(\overline{z}))\dd x(z)} \nonumber.
\eeq
\medskip

\item[$\bullet$] For $k + 2g - 2 > 0$, we define recursively the $k$-forms $\mathcal{W}_k^{(g)}(z_1,\ldots,z_k)$ by:
\bea\label{toprec}
 \mathcal{W}_k^{(g)}(z_1,\underbrace{z_2,\ldots{},z_k}_K)
&=& \sum_i\Res_{z \rightarrow a_i}K(z_1,z)\Big[\mathcal{W}_{k + 1}^{(g-1)}(z,\overline{z},K) \cr
&&  + \sum_{J \subseteq K,\; 0 \leq h \leq g}'\mathcal{W}_{|J| + 1}^{(h)}(z,J)\mathcal{W}_{k - |J|}^{(g -
 h)}(\overline{z},K\setminus J)\Big],
\eea
where $\sum'$ ranges over $(J,h) \neq (\emptyset,0),(I,g)$.
\medskip

\item[$\bullet$] $\mathcal{W}_0^{(g)} = \mathcal{F}_g$ are defined for $g \geq 2$ by
\beq
 \mathcal{F}_g = \frac{1}{2 - 2g}\sum_i\Res_{z
\rightarrow a_i}\left(\mathcal{W}_1^{(g)}(z)\Phi(z)\right), \nonumber \eeq
where $\Phi$
is a primitive of $y\dd x$ locally at the $a_i$'s, i.e. $d\Phi=y dx$. 
\medskip

\item[$\bullet$]
The definition of $\mathcal{F}_1$ and $\mathcal{F}_0$ is more involved, and we refer the reader to  \cite{EOFg}.
\end{itemize}

\bigskip

\subsection{Main properties of the ${\mathcal W}_k^{(g)}$}

\begin{itemize}
\item \underline{Symmetry}\phantom{e} $\forall k,g, \quad \mathcal{W}_k^{(g)}(z_1,\ldots,z_k)$ is symmetric in
$z_1,\ldots,z_k$. 
\medskip

\item \underline{Invariance}\phantom{e} $\forall k,g, 2g + k - 2 > 0, \quad \mathcal{W}_k^{(g)}$ is
unchanged if we add to $y$ a rational function of $x$. 
\medskip

\item \underline{Exchange invariance}
\phantom{e} $\forall g\geq 2 \quad \mathcal{F}_g$ is unchanged if we exchange $x$ and $y$. 
\medskip

\item \underline{Deformation}\phantom{e} If we perform an infinitesimal
deformation $(\delta x,\delta y)$ of the spectral curve, the
$\mathcal{W}_k^{(g)}$'s change. Let us introduce
$$
\Omega(z) = \delta x(z)\dd y(z) - \delta y(z) \dd x(z).
$$
This form does not depend on the parametrization $z$, and can always be represented as $\int_{z' \in \gamma}B(z,z')\Lambda(z')$ for some path $\gamma$ and
some meromorphic function $\Lambda$ defined in its neighborhood (this data is called the dual of $\Omega$). Then
\beq
\label{eq:Deform}\delta \mathcal{W}_k^{(g)}(z_1,\ldots{},z_k) = \int_{z \in \gamma} \Lambda(z)\mathcal{W}_{k +
1}^{(g)}(z,z_1,\ldots{},z_k).
\eeq 
\medskip

\item \underline{Limits}\phantom{e} $W_k^{(g)}(\mathcal{S})$ is compatible with limits of curves. 
\medskip

\item \underline{Link to matrix models}\phantom{e} It was proved in \cite{EOFg} that, if our matrix model has a topological expansion property:
$$
W_k(z_1,\dots,z_k) = \left<\prod_{i=1}^k\tr\frac{1}{x(z_i) - M}\right>_c \nonumber = \sum_{g = 0}^{\infty} g_s^{2g-2 + k}W_k^{(g)}(z_I),
$$
and
$$
\ln Z = \sum_{g=0}^\infty g_s^{2g-2}\, F_g,
$$
then, loop equations imply that:
$$
W_k^{(g)}(z_1,\dots,z_k)dx(z_1)\dots dx(z_k) = \mathcal{W}_k^{(g)}(z_1,\dots,z_k)
$$
where $\mathcal{W}_k^{(g)}$ are computed using the spectral curve ${\mathcal S}=(x(z),y(z))$ presented in Section~\ref{sec:SCcar} (except for $(k,g) = (1,0), (2,0)$, which receive simple additional contributions). Similarly, for $g\geq 2$:
$$
F_g = \mathcal{F}_g.
$$
($F_0$ and $F_1$ also receive simple corrections).
\end{itemize}
\medskip

\noindent Since the $\mathcal{F}_{g}(\mathcal{S})$'s are invariant under
transformations of $\mathcal{S}$ which leave $|\dd x\wedge \dd y|$ unchanged,
we call them \emph{symplectic invariants} of $\mathcal{S}$.

\medskip

In the following section we shall apply the general theory of \cite{EOFg} to our Hurwitz matrix integral Eqn~\ref{eq:ZZ}.


\section{Spectral curve of the Hurwitz matrix model}
\label{sec:SpectralCurve}


In our case, the spectral curve of our matrix model must be
determined order by order in powers of $g_s$ and $t$. Here, we only give the
result. The proof is quite technical,  and is deferred to Appendix~\ref{app:SpectralDerivation}.
It relies on the fact that, after a suitable shift, to leading order in $g_s$ and $t$, we have a
Gaussian matrix integral. In particular, this implies that the
spectral curve has genus 0, i.e. it can be parametrized with a
uniformization variable $z\in \mathbb C$.
By composition with some homographic
map, we can choose a parametrization for which $x(z)\sim z$ when $z \rightarrow \infty$ and $x(0)=0$.

Since our potential $V$ is of the form Eqn. \ref{eq:typicalV} with $n\to\infty$, our spectral curve is of the form Eqn. \ref{genspcurve}.
Therefore we guess that the spectral curve must be of the form:
\beq \label{eq:SC1}
\mathcal{S}(\p;t) :\quad
\left\{\begin{array}{l}
x(z)= z + g_s\sum_{i=1}^N \frac{1}{(z-z_i)y_i}+{1\over z_i y_i} \\
y(z)= - z  + \ln{(z/t)} + c_0 + {c_1\over z} - \sum_{l=1}^\infty {B_{2l}\, g_s^{2l}\over 2l}\, \left(f_{2l}(z)-{f_{2l,1}\over z}\right)
\end{array}\right.
\eeq
where
$$
\left\{\begin{array}{l}
y(z_i)=R_i=\ln v_i \\
y_i=y'(z_i) \\
\end{array}\right.
$$
and where
$$
f_l(z) = \Res_{z'\to 0} {dz'\over z-z'}\,\, (x(z'))^{-l} = \sum_{j=1}^l f_{l,j}\, z^{-j}
$$
is such that $x(z)^{-l}-f_l(z)$ has a finite limit when $z\to 0$.
$c_0$ is chosen such that  $V'(x(z))-y(z)\sim O(1/z)$ at large $z$:
$$
c_0=(N-\frac{1}{2})\,g_s + g_s\sum_{i=1}^N {1\over z_i\,y_i}.
$$
The coefficient $c_1$ is
$$
c_1={g_s\over 2} - \sum_{l=1}^\infty {B_{2l}\, g_s^{2l}\over 2l}\, f_{2l,1},
$$
and one can check (this comes from the fact that the sum of all residues of $V'(x)$ must vanish) that it is such that $V'(x(z))-y(z)\sim Ng_s/z$ at large $z$ i.e.:
$$
c_1=(N-\frac{1}{2})\,g_s - g_s\sum_{i=1}^N {1-z_i\over z_i\,y_i}.
$$

Each term $z_i, y_i, c_0, c_1$, is to be viewed as a power series in $g_s$ and in $t$.
To the first few orders in $g_s$ we have:
$$
z_i = L(tv_i) + {g_s\over 1-L(tv_i)}\left({1+L(tv_i)\over 2} + L(tv_i)\sum_j {L(tv_j)\over 1-L(tv_j)}\right) + O(g_s^2)
$$
$$
c_0 = g_s\left(-{1\over 2} - \sum_{i=1}^N {L(tv_i)\over 1-L(tv_i)}\right) + O(g_s^2)
$$
$$
c_1 =  {- g_s\over 2} + O(g_s^2)
$$
and notice that $L(tv_i)$ is a power series in $t$:
$$
L(tv_i) = \sum_{m=1}^\infty \frac{m^{m-1}\,t^m\, v_i^m}{m!}.
$$

The proof that $\mathcal{S}(\p;t)$ is the correct spectral curve for our problem is given in Appendix~\ref{app:SpectralDerivation}. It is obtained by computing the spectral curve order by order in the small $g_s$ and $t$ expansion.

\medskip
The computation of ${\mathcal W}_k^{(g)}$'s in the topological recursion formula Eqn~\ref{toprec} involve taking residues at all zeroes of $x'(z)$, i.e. involves symmetric rational functions of the $z_i$'s, and one can see that the coefficients of ${\mathcal W}_k^{(g)}$ are, order by order in $g_s$ and $t$, polynomials of $p_m=g_s\sum_i v_i^m$, as required for the computation of Hurwitz numbers indeed.

\subsection{Spectral curve at $g_s=0$}

At $g_s=0$, the spectral curve reduces to:
\beq \label{eq:SC0}
\mathcal{S}_0 :\quad
\left\{\begin{array}{l}
x(z)= z  \\
y(z)= - z  + \ln{(z/t)}
\end{array}\right.
\eeq
i.e. $x=L(t e^y)$, where $L$ is the Lambert function.
Up to exchanging the roles of $x$ and $y$, this is the Lambert curve $\mathcal{S}_{\mathrm{Lambert}}$ appearing in Bouchard-Mari\~no \cite{BM}.

\subsection{The symplectic invariants}

So far, from the general theory of matrix models and from general properties of the topological recursion, we have proved that the generating function of simple
Hurwitz numbers of genus $g$ is the symplectic invariant
\beq
\label{eq:Fg}
F_g(\p;t) = \sum_{n} t^n\, \sum_{|\mathbf{\mu}|=n}
\frac{p_{\mathbf{\mu}}}{(2g - 2 + n + \ell(\mathbf{\mu}))!}\,
H_{g,\mathbf{\mu}} = {\mathcal F}_g({\mathcal S}(\p;t)),
\eeq
where
${\mathcal S}(\p;t)$ is the spectral curve of Eqn~\ref{eq:SC1} (in fact we have proved it only for $g\geq 2$, and we consider that the cases $g=0$ and $g=1$ are easier).

\medskip
The computation of symplectic invariants ${\mathcal F}_g$ involves computing residues at the zeroes of $x'(z)$, and there are $2N$ such zeroes, which makes the computation complicated.

One may use the invariance properties of ${\mathcal F}_g$, under the exchange of $x$ and $y$.
Indeed, the zeroes of $y'(z)$ are much simpler to compute, order by order in powers of $g_s$.
At $g_s=0$, $y'(z)$ has only one zero located at $z=1$.

Therefore, we introduce a new spectral curve satisfying
$$
F_g(\p;t)  = \mathcal{F}_g(\widetilde{\mathcal{S}}(\p;t)),
$$
defined by exchanging $x$ and $y$:
$$
\boxed{\widetilde{\mathcal{S}}(\p,g_s;t) :
\left\{\begin{array}{l}
x(z)= - z  + \ln{(z/t)} + c_0 + {c_1\over z} - \sum_{l=1}^\infty {B_{2l}\, g_s^{2l}\over 2l}\, f_{2l}(z)\\
y(z)= z + g_s\sum_{i=1}^N \frac{1}{(z-z_i)y_i}+{1\over z_i y_i}
\end{array}\right. ,}
$$
where now $x(z_i)=\ln v_i$ and $y_i=x'(z_i)$.

\subsection{The correlation forms}

So far, we have introduced the $F_g$'s  as generating functions which encode the simple Hurwitz numbers $H_{g,\mu}$ by expansion on a proper basis of polynomials of the $p_m$'s.
Bouchard and Mari\~no \cite{BM} define another generating function, namely the function of $k$-variables:
$$
H^{(g)}(x_1,\ldots,x_k) = \sum_{\ell(\mathbf{\mu}) = k}\, t^{|\mu|}\,\, \frac{\prod_{i = 1}^k \mu_i\cdot M_{\mathbf{\mu}}(x_1,\ldots,x_k)}{(2g - 2 + |\mathbf{\mu}| + k)!}H_{g,{\mathbf{\mu}}},
$$
where $M_{\mathbf{\mu}}(\mathbf{x}) = \sum_{\sigma \in
\mathfrak{S}_k} \prod_{i = 1}^{k} x_{\sigma(i)}^{\mu_i}$ are the (un-normalized) symmetric monomials.

It is easy to relate both. If we recall the combinatorial definition Eqn~\ref{eq:Fg}
$$
F_g(\p;t) = \sum_{\mu} t^{|\mathbf{\mu}|}\,\, \frac{p_{\mathbf{\mu}}}{(2g - 2 + n + \ell(\mathbf{\mu}))!}\, H_{g,\mathbf{\mu}},
$$
we see that
$$
H^{(g)}(v_1,\dots,v_k)
= \left.\left(\frac{v_1\dots v_k}{g_s^{k}}\frac{\partial^k F_g}{\partial v_1 \dots \partial v_k}\right)\right|_{g_s = 0}
= \left.\left(\frac{1}{g_s^{k}}\frac{\partial^k F_g}{\partial R_1 \dots \partial R_k}\right)\right|_{g_s = 0}.
$$
We recall that $R_i=\ln v_i$.

The deformation property Eqn~\ref{eq:Deform} of symplectic invariants allows us to calculate their derivatives.
When we perform an infinitesimal variation $v_i \rightarrow v_i + \delta v_i$, i.e. a variation $R_i \rightarrow R_i + \delta R_i$ on the spectral curve, we need to compute
$$
\Omega_i(z) = \delta x(z) \dd y(z) - \delta y(z) \dd x(z).
$$

First notice that the form $y\dd x$ is a meromorphic form (its singularities are poles).
It has simple poles of constant residues $g_s$ near $z=z_j$, i.e. locally near $z_j$ we have
$$
y \sim \frac{g_s}{x-R_j},
$$
which implies that, locally near $z_j$,
$$
\Omega_i = -{g_s\,\delta_{i,j}\delta R_j}\,\, \frac{\dd x}{(x-R_j)^2} + O(1)
$$
Then, observe that the other poles of $y\dd x$ are independent of $R_i$. For example, $(V'(x)-y)\dd x$ has a simple pole at $\infty$, with residue $Ng_s$ independent of $R_i$. Therefore, $\Omega_i$ has no residue, and thus no pole at $\infty$. Similarly there is no pole at $z=0$. The final result is that $\Omega_i$ is a meromorphic form with poles only at $z_i$:
$$
\frac{1}{g_s}\,\, \Omega_i(z) = - \frac{\delta R_i\, \mathrm{d}z}{(z-z_i)^2\,y_i}.
$$

\noindent $\Omega_i(z)$ can be written in term of the Bergman kernel $B(z,z')=\frac{\dd z\,\dd z'}{(z-z')^2}$ as
$$
\Omega_i(z) = -g_s\,\delta R_i\, \Res_{z'\to z_i} \, B(z,z')\, \frac{1}{(z'-z_i)\,y_i}.
$$

Then, the general theorems about the ${\mathcal F}_g$'s tell us that
\bea
\delta {\mathcal F}_g
&=& g_s\delta R_i\,\Res_{z' \rightarrow z_i}\,\mathcal{W}_1^{(g)}(z')\,\,\frac{1}{(z'-z_i)\,y_i}
\nonumber \\
&=& g_s\delta R_i\,\frac{\mathcal{W}_1^{(g)}(z_i)}{\dd{}x(z_i)},  \nonumber
\eea
and more generally
\beq
\delta \mathcal{W}_k^{(g)}(z_1,\ldots{},z_k) = g_s \delta R_i\,\,\Res_{z' \rightarrow z_i} \,\mathcal{W}_{k+1}^{(g)}(z_1,\ldots,z_k,z')\left[\frac{1}{(z'-z_i)y_i}\right]. \nonumber
\eeq
The result is that
$$
\frac{1}{g_s^k}\frac{\partial^k \mathcal{F}_g}{\partial v_1 \dots \partial v_k}= \left.  \left(
  \frac{\mathcal{W}_k^{(g)}(z_1,\dots,z_k)}{\dd x(z_1)\cdots\dd x(z_k)}.
  \right)  \right|_{R_1 = x(z_1),\ldots,R_k=x(z_k)}
$$

As a final step,  we take the limit $g_s \rightarrow 0$.
The limit of the spectral curve $\widetilde{\mathcal S}({\mathbf v};t)$ is simply
$$
{\mathcal S}_{\rm Lambert} \,\, : \,\, \left\{\begin{array}{l}
x(z)= - z + \ln{(z/t)}  \\
y(z)= z,  \\
\end{array}\right. \nonumber
$$
i.e. it is the Lambert spectral curve: $y = L(t e^x)$.

In other words, we have proved that the function $H^{(g)}(v_1,\dots,v_k)$ is the correlation form $\mathcal{W}_k^{(g)}$ of the Lambert spectral curve ${{\mathcal{S}}_{\rm Lambert}}$:
$$
\boxed{H^{(g)}(v_1,\dots,v_k) = \left[\frac{{\mathcal{W}_k^{(g)}}(z_1,\dots,z_k)}{\dd x(z_1)\cdots\dd x(z_k)}\right]({\mathcal{S}}_{\rm Lambert}),}
$$
where $x(z_i)=\ln v_i=R_i$, i.e. $z_i=L(t v_i)$.

\smallskip
This is precisely the Bouchard-Mari\~no conjecture.

\section{Relationship with intersection numbers, Kontsevich integral and \textsc{elsv} formula}
\label{sec:ELSV}

Notice that the Lambert spectral curve
$$
{\mathcal S}_{\rm Lambert} \,\, : \,\, \left\{\begin{array}{l}
x(z)= - z + \ln{(z/t)}  \\
y(z)= z  \\
\end{array}\right. \nonumber
$$
has only one branchpoint (solution of $x'(z)=0$), given by $z=1$.
This is the reason why the Bouchard-Mari\~no conjecture is so efficient to compute Hurwitz numbers.

Since the topological recursion for computing the $\mathcal{W}_k^{(g)}$'s and ${\mathcal F}_g$'s, involves only the computation of residues at the branch point, we may perform a Taylor expansion near $z=1$ :
Let us define:
$$y=1+\zeta$$
and
$$
\xi^2 = -2 (x+1+\ln t)
$$
We have, in the limit $\zeta\to 0$:
$$
\frac{1}{2}\xi^2 = \frac{\zeta^2}{2} -\frac{\zeta^3}{3} + \frac{\zeta^4}{4} + \dots = \sum_{m\geq 2} \frac{(-1)^m\,\zeta^m}{m}
$$
and we invert that expansion $y=1+\xi + \frac{\xi^2}{3} + \frac{\xi^3}{36}- \frac{\xi^4}{270}+\frac{\xi^5}{6\, . 6!}+\cdots $,
which we write:
$$
y = 1-2 \xi +\sum_{m\geq 1} t_{m+2} \xi^m
$$
In other words, the $\mathcal{W}_k^{(g)}$'s and ${\mathcal F}_g$'s of the Lambert curve ${\mathcal{S}}_{\rm Lambert}$, are the same as the $\mathcal{W}_k^{(g)}$'s and ${\mathcal F}_g$'s of the following spectral curve ${\mathcal{S}}_K$:
$$
{\mathcal S}_K \,\, : \,\, \left\{\begin{array}{l}
x(\xi)= -1-\ln t -\frac{1}{2}\,\xi^2  \\
y(\xi)= 1-2 \xi +\sum_{m\geq 1} t_{m+2} \xi^m  \\
\end{array}\right. \nonumber
$$
This spectral curve is exactly the Kontsevich spectral curve for times $t_m$'s (see \cite{EKM}).
Here the $t_m$'s satisfy the following recursion $t_2=0$, $t_3=3$, $t_4=\frac{1}{3}$, and for $m\geq 4$:
\begin{equation}
\label{eq:TimeParameters}
t_{m+1} = \frac{t_{m}}{m} -  \frac{1}{2}\,\sum_{l=2}^{m-2}\, t_{l+2}\,t_{m+2-l}
\end{equation}
We form the following series:
$$
f(z) = \sum_{m=1}^\infty \frac{(2m+1)!}{m!}\,\frac{t_{2m+3}}{2-t_3}\, z^m
$$
and
$$
g(z) = -\ln{(1-f(z))} = \sum_{m=1}^\infty \tilde t_m\, z^m
$$
we find to the first orders:
$$
g(z) = -\frac{z}{6} + \frac{z^3}{45} - \frac{ 8 z^5}{315} + \frac{ 8 z^7}{105} + \cdots
$$
Then, it was found in \cite{EKM} that:
\bea
\nonumber \mathcal{W}_k^{(g)}(z_1,\ldots{},z_k)
&=& \frac{1}{2^{3g-3+k}}\, \sum_{d_0+\dots+d_k=3g-3+k} \sum_{j=1}^{d_0} \frac{1}{j!} \,\, \sum_{m_1+\dots+m_j=d_0, m_i>0} \\
\nonumber  &&  \prod_{i=1}^k \frac{(2d_i+1)!}{d_i!}\,\frac{\dd{}z_i}{z_i^{2d_i+2}}\,\prod_{i=1}^j \tilde t_{m_i} \,\, \left< \prod_{i=1}^j \kappa_{m_i}\, \prod_{i=1}^k \psi_i^{d_i} \right>_{\overline{\mathcal M}_{g,k}}
\eea
where $\overline{\mathcal M}_{g,k}$ is the stable compact moduli space of Riemann surfaces of genus $g$ with $k$ marked points, and $\kappa_j$ is the $j^{\rm th}$ Mumford's tautological class, and $\psi_i=c_1({\mathcal L}_i)$ is the first Chern class of the cotangent bundle at the $i^{\rm th}$ marked point.

In other words, just by looking at the Lambert spectral curve, we see that there is a relationship between the generating function for Hurwitz numbers of genus $g$ with a monodromy of length $k$, and the generating function for intersection numbers of tautological classes on $\overline{\mathcal M}_{g,k}$.

This type of relationship is completely natural and expected. The link coming from the \textsc{elsv} formula
\cite{ELSV}, relating Hurwitz numbers to Hodge integrals:
\begin{equation*}
 H_{g, \mu} = \frac{(2g-2 + \ell(\mu) + |\mu|)!}{|\Aut \mathbf{\mu}|} \prod_{i=1}^{\ell(\mathbf{\mu})}\frac{\mu_i^{\mu_i}}{\mu_i!}
  \int_{\overline{\mathcal M}_{g, \ell(\mathbf{\mu})}} \frac{\Lambda_g^{\vee}(1)}{\prod_{i=1}^{\ell(\mathbf{\mu})}(1 - \mu_i \psi_i)},
\end{equation*}
where $\Lambda ^{\vee} _g(t) = \sum_{i=0}^g \lambda_i t^i$ is the total Chern class of the Hodge bundle $\mathbb{E}$ over $\overline{\mathcal M}_{g,n}$.
In fact, the change of variables taking the Hurwitz generating function to the generating function of all Hodge integrals with a single $\lambda$ factor is given by the Lambert curve itself \cite{BM, GJV, Kazarian}. Furthermore, Mumford's formula \cite{FP, Mum}
\begin{equation*}
 \ch(\mathbb{E}) = g + \sum_{l=1}^{\infty} \frac{B_{2l}}{(2l)!} \left(
  \kappa_{2l-1} + \frac{1}{2}\iota_{*} \sum_{i=0}^{2l-2} (-1)^i \psi^i \bar{\psi}^{2l-2-i}
 \right)
\end{equation*}
allows one to express Hodge integrals in terms of $\psi$ class intersections. The time parameters $t_m$ appearing in Eqn~\ref{eq:TimeParameters} are closely related to these topics. Mironov and Morozov \cite{MirMor} have previously considered similar constructions.

An other natural question arising is the relationship between the Bouchard--Mari\~no conjecture and the cut--and--join equation \cite{GJ}. Since they are both recursive algorithms for computing Hurwitz numbers, it seems likely that they should be related. In fact, they are, appropriately interpreted, completely equivalent. This topic, as well as a more detailed discussion of the deformation of the Kontsevich model to the Hurwitz model, are deferred to future papers.


\section{Conclusion}\label{sec:conclusion}

With the integral representation of $\mathrm{U}(N)$ characters, it is possible to express in general the partition function $Z$ of Hurwitz numbers as a matrix model. In the case of simple Hurwitz numbers, we obtain a 1-matrix model with external field, whose spectral curve is found by solving the master loop equation:
$$
\widetilde{\mathcal{S}}(\p;t) :
\left\{\begin{array}{l}
x(z)= - z  + \ln{(z/t)} + c_0 + {c_1\over z} - \sum_{n=1}^\infty {B_{2n}\, g_s^{2n}\over 2n}\, f_{2n}(z)\\
y(z)= z + g_s\sum_{i=1}^N \frac{1}{(z-z_i)y_i}+{1\over z_i y_i}
\end{array}\right. .
$$
Our main results are:
\bea
F_g(\p;t) & = & \mathcal{F}_g(\widetilde{\mathcal{S}}(\p;t)) \nonumber\\
H^{(g)}(v_1,\ldots,v_k) & = & \left[\frac{\mathcal{W}_k^{(g)}(z_1,\ldots,z_k)}{\mathrm{d}x(z_1)\cdots\mathrm{d}x(z_k)}\right](\widetilde{\mathcal{S}}_{\textrm{Lambert}}) \qquad \textrm{where}\,\left\{\begin{array}{l} x(z_i) = \ln v_i \\ y(z_i) = L(tv_i) \end{array}\right. \nonumber
\eea

It provides an algorithm, namely the topological recursion of matrix models, to compute the $H_{g,\mathbf{\mu}}$ by the residue formula of Paragraph~\ref{sec:Toporec}, with only one branchpoint involved. As a matter of fact, this recursion relation between simple Hurwitz numbers is understood to be equivalent to the Laplace transform of the cut-and-join equation with help of the \textsc{elsv} formula. Besides, $Z$ for simple Hurwitz numbers is the time evolution of the \textsc{kp} $\tau$-function, a fact agreeing with its one-matrix model representation. We also see explicitly on the Lambert curve $\mathcal{S}_{\textrm{Lambert}} = \lim_{g_s \rightarrow 0} \widetilde{\mathcal{S}}(\p;t)$ the relation between $Z$ and the Kontsevich $\tau$-function.

We hope that our matrix model-minded methods could help investigating double Hurwitz numbers (where $Z$ is a Toda $\tau$-function) and further.


\section*{Acknowledgments}
We would like to thank I. Kostov, M.~Mari\~ no and N.~Orantin for useful and
fruitful discussions on this subject. The work of B.E. is partly
supported by the Enigma European network \textsc{mrt-ct-2004-5652}, by the
\textsc{anr} project G\'eom\'etrie et int\'egrabilit\'e en physique
math\'ematique \textsc{anr-05-blan-0029-01}, by the \textsc{anr} GranMa  grant
\textsc{anr-08-blan-0311-03}, by the European Science Foundation through the
Misgam program, by the Quebec government with the \textsc{fqnrt}.


\section{Appendix: proof of the spectral curve}
\label{app:SpectralDerivation}

The proof works order by order in $g_s$ and $t$, and it relies on the fact
that, to leading order, we have a Gaussian matrix integral.

\subsection{Shift of the matrix model}

We start from:
$$
Z(\p, g_s;t)= \frac{g_s^{-N^2}}{N!}\frac{\Delta(\R)}{\Delta(\v)} \int_{\Hermitian_N (\mathcal{C})} dM e^{-\frac{1}{g_s} \tr \bigl[ V(M)-M\R \bigr]},
$$
where the potential $V(x)$ is:
$$
V(x) = - \frac{x^2}{2} +  g_s(N-\frac{1}{2}) x + \left(\ln (g_s/t) +
i\pi \right)x - g_s\ln{\bigl(\Gamma(-x/g_s )\bigr)} + C_t,
$$
and $C_t$ does not depend on $x$. Order by order in $g_s$ we have the Stirling expansion:
$$
V'(x) = -x + \ln{(x/t)} + g_s(N-\frac{1}{2}) + {g_s\over 2x} - \sum_{l = 1}^\infty {B_{2l}\,g_s^{2l}\over 2l\, x^{2l}}.
$$

We need to compute this matrix integral in the small $g_s$ and $t$ expansion (up to a constant factoring out of the integral).

First let us perform a shift
$$
\widetilde{M} = M - \tilde{\mathbf R}
$$
where $\tilde{\mathbf R}={\rm diag}(\tilde R_1,\dots,\tilde R_N)$ is such that:
$$
V'(\tilde R_i)=R_i.
$$
The equation $V'(\tilde R_i)=R_i$ has several solutions, we choose the one which is a power series in $g_s$ and $t$:
$$
\tilde R_i = L(t v_i) - g_s({1\over 2} + {NL(t v_i)\over 1-L(t v_i)}) + \dots = \sum_{l \geq 0} g_s^l \tilde R_{i,l}(v_i).
$$
and we choose the determination of the Lambert function such that $L(tv_i)$ has a small $t$ expansion $L(tv_i) = tv_i + t^2 v_i^2 + \dots = \sum_{m \geq 1} m^{m-1} (tv_i)^m/m!$.

\medskip
Then we have:
$$
Z(\p, g_s;t)= \frac{g_s^{-N^2}}{N!}\frac{\Delta(\R)}{\Delta(\v)}\, e^{-\frac{1}{g_s} \tr \bigl[ V(\tilde{\mathbf R})-\tilde{\mathbf R}\R \bigr]} \int_{\Hermitian_N (\mathcal{C})} d\widetilde M\, e^{-\frac{1}{g_s} \tilde {\mathcal V}(\widetilde M)},
$$
where we decompose $\tilde {\mathcal V}(\widetilde M)$
$$
\tilde {\mathcal V}(\widetilde M) = \tilde {\mathcal V}_2(\widetilde M) + \tilde {\mathcal V}_{\ge 3}(\widetilde M)
$$
into a quadratic function of $\widetilde M$
$$
\tilde {\mathcal V}_2(\widetilde M) = {1\over 2} \tr \left(  -\widetilde M^2 + g_s \sum_{j=0}^\infty  {1\over \tilde{\mathbf R}-j g_s}  \widetilde M {1\over \tilde{\mathbf R}-j g_s} \widetilde M \right)
$$
and $\tilde {\mathcal V}_{\geq 3}$ contains all the higher degree terms (we will not need explicit expressions, however, the interested reader can derive them easily).

Notice that $\tilde {\mathcal V}_2(\widetilde M)$ and $\tilde {\mathcal V}_{\geq 3}(\widetilde{M})$ have a small $g_s$ expansion (for instance approximate the sum $\sum_j$ in $\tilde {\mathcal V}_2$ by a Riemann integral).

\medskip

Order by order in $g_s$ we have:
\bea
Z(\p, g_s;t)
&=& \frac{g_s^{-N^2}}{N!}\frac{\Delta(\R)}{\Delta(\v)}\, e^{-\frac{1}{g_s} \tr \bigl[ V(\tilde{\mathbf R})-\tilde{\mathbf R}\R \bigr]}  \nonumber \cr
&& \quad \int_{\Hermitian_N (\mathcal{C})} \mathrm{d}\widetilde M\, e^{-\frac{1}{g_s} \tilde {\mathcal V}_2(\widetilde M)}\,\, \sum_{m=0}^\infty {(-1)^m\over g_s^m\,\,m!} (\tilde {\mathcal V}_{\geq 3}(\widetilde M))^m, \nonumber 
\eea
If we rescale $\widetilde M=\sqrt{g_s}\,A$, we have
\bea
Z(\p, g_s;t)
&=& \frac{1}{N!}\frac{\Delta(\R)}{\Delta(\v)}\, e^{-\frac{1}{g_s} \tr \bigl[ V(\tilde{\mathbf R})-\tilde{\mathbf R}\R \bigr]} \nonumber \cr 
&& \quad \int_{\Hermitian_N (\mathcal{C})} \mathrm{d}A\, e^{- \tilde {\mathcal V}_2(A)}\,\, \sum_{m=0}^\infty {(-1)^m\over g_s^m\,\,m!} (\tilde {\mathcal V}_{\geq 3}(A/\sqrt{g_s}))^m. \nonumber 
\eea
Since $\tilde {\mathcal V}_{\geq 3}(A/\sqrt{g_s}) = O(\sqrt{g_s})$, we see that order by order in powers of $g_s$, we may exchange the sum and integral. Therefore:
\bea
Z(\p, g_s;t)
&=& \frac{g_s^{-N^2}}{N!}\frac{\Delta(\R)}{\Delta(\v)}\, e^{-\frac{1}{g_s} \tr \bigl[ V(\tilde{\mathbf R})-\tilde{\mathbf R}\R \bigr]}  \nonumber \cr
&& \quad \sum_{m=0}^\infty {(-1)^m\over g_s^m\,\,m!}\int_{\Hermitian_N (\mathcal{C})} \mathrm{d}\widetilde M\, e^{-\frac{1}{g_s} \tilde {\mathcal V}_2(\widetilde M)}\,\,  (\tilde {\mathcal V}_{\geq 3}(\widetilde M))^m, \nonumber 
\eea

More generally, expectation values of polynomials $Q_p(\widetilde M)$ are computed as formal power series, whose coefficients are polynomial moments of a gaussian integral:
\bea
< Q_p(\widetilde M)> &=& \frac{\sum_{m=0}^\infty {(-1)^m\over g_s^m\,\,m!} \int_{\Hermitian_N (\mathcal{C})} \mathrm{d}\widetilde M\, e^{-\frac{1}{g_s} \tilde {\mathcal V}_2(\widetilde M)}\,\, (\tilde {\mathcal V}_{\geq 3}(\widetilde M))^m\,\, Q_p(\widetilde M)}
{\sum_{m=0}^\infty {(-1)^m\over g_s^m\,\,m!} \int_{\Hermitian_N (\mathcal{C})} \mathrm{d}\widetilde M\, e^{-\frac{1}{g_s} \tilde {\mathcal V}_2(\widetilde M)}\,\, (\tilde {\mathcal V}_{\geq 3}(\widetilde M))^m}. \nonumber
\eea
In this form, we can use Wick's theorem.
It shows that, if $Q_p(\widetilde M)$ is any homogeneous polynomial of total degree $p$ in the entries of
the matrix $\widetilde M$, the expectation value $\left<
Q_p(\widetilde{M}-\widetilde{\R}) \right>$ is a power series
in $g_s$ and $t$. It is expressed as a finite sum of
connected\footnote{Connected because the normalization factor
$\widetilde{Z}^{-1}$ is included in the expectation value.}
fat-graphs. When we restrict the sum to the fatgraphs of
genus $g$, we note it with a superscript $^{(g)}$. We claim that,
for our matrix model, this expectation value is $O(g_s^{p/2})$, and
$O(g_s^{2g - 1})$. It can be seen as follows : fatgraphs
contributing to the sum have one vertex of degree $p$, and $v$ internal
vertices of degree $\geq 3$ coming from $(\tilde {\mathcal V}_{\geq 3})^v$. If we call $f$ the
number of faces, and $e$ the number of edges in the fatgraph, its
Euler characteristic is $2-2g=(v+1) - e + f$. The power of $g_s$
coming from the gaussian integral is $j \geq e - v$ (indeed, we have $g_s^e$ coming from the gaussian integral with $e^{-\frac{1}{g_s}\tilde {\mathcal V}_{2}(\widetilde M)}$, and $g_s^{-v}$ accompanying $(\tilde {\mathcal V}_{\geq 3})^v$, and in addition $\tilde {\mathcal V}_2$ and $\tilde {\mathcal V}_{\geq 3}$ themselves have a $g_s$ expansion).
So, we have:
$$
j\geq e - v  = 2g - 1 + f \geq 2g-1.
$$
On the other hand, the number of half-edges is $2e = p +\sum_i i
n_i$ where $n_i$ is the number of internal vertices of degree $i
\geq 2$, and we have $v=\sum_i n_i$. So:
$$
j\geq e - v = \sum_i (\frac{i}{2} - 1)n_i + \frac{p}{2} \geq
\frac{p}{2}.
$$

In particular, let us show how to define the $W_1^{(g)}$'s, the
topological expansion of the one-point correlation function.
Consider :
$$
T_{p}(x) = \left<\tr
\left[\left(\frac{1}{x-\widetilde{\R}}\,\widetilde{M}\right)^p\frac{1}{x-\widetilde{\R}}\right]
\right>.
$$
It is a double power series, whose coefficients are rational
functions of $x$, and are polynomial gaussian expectation values of $\widetilde{M}$. In its representation as a sum over fatgraphs, we
collect those of genus $g$ to define $T_p^{(g)}(x)$. Its
coefficients are still rational functions of $x$. Since $T_p^{(g)}(x)=O(g_s^{p/2})$, we write
$$
T_p^{(g)}(x)=\sum_{j\geq p/2} g_s^j\,\, T_{p,j}^{(g)}(x),
$$
and thus we have, in the sense of formal power series of $g_s$:
$$
\sum_{p=0}^{\infty} T_p^{(g)}(x) =\sum_{j\geq 2g-1} g_s^j\,\, \sum_{p=0}^{2j} T_{p,j}^{(g)}(x).
$$
We are in position to define $W_1^{(g)}(x)$ as:
$$
g_s^{2g-1}\,W_1^{(g)}(x) = \sum_{j = 2g-1}^\infty g_s^j\sum_{p=0}^{2j} T_{p,j}^{(g)}(x).
$$
$W_1^{(g)}(x)$ is thus a formal power series in powers of $g_s$ and $t$, whose coefficients are rational functions of $x$.
These definitions give a meaning to the equality between formal double power series:
$$
W_1(x) = \sum_{g=0}^\infty g_s^{2g-1}W_1^{(g)}(x),
$$
where $W_1(x)$ is the resolvent :
$$
W_1(x) \mathop{=}^{\mathrm{formal}} \,\, \left< \tr
\frac{1}{x-M} \right>.
$$

In a similar manner, one can define $g_s^{2g-2+k}\,
W^{(g)}_k(x_1,\dots,x_k)$ as the formal double power series
computing the sum over ($c$ for connected) fatgraphs  of genus $g$
arising in the correlation function:
$$
W_k(x_1,\dots,x_k) \mathop{=}^{\mathrm{formal}} \,\, \left<\prod_{i
= 1}^{k} \tr\left(\frac{1}{x_i-M}\right)\right>_c.
$$
By construction we have, in the sense of formal series:
$$
W_k(x_1,\dots,x_k) = \sum_{g = 0}^\infty g_s^{2g - 2 + k}\,
W^{(g)}_k(x_1,\dots,x_k).
$$

To sum things up, the correlation functions
$W^{(g)}_k(x_1,\dots,x_k)$ can be defined as formal power series in  $g_s$ and $t$, such that the coefficients are rational functions of the $x_i$'s.
It is defined by collecting the fatgraphs of genus $g$ in the Wick theorem's expansion of gaussian integrals.

\medskip

The loop equations of gaussian matrix integrals are well known, and they imply that the $W_k^{(g)}$ satisfy the topological recursion of \cite{EOFg}.

\medskip

For $W_1^{(0)}$, or more precisely for $Y(x)= V'(x)-W_1^{(0)}(x)$, the loop equations read (see \cite{EOFg}):
\bea\label{loopeqapp}
V'(x)-Y
&=& \left<\tr {V'(x)-V'(M)\over x-M}\, {1\over Y-\mathbf R}\right>^{(0)} \nonumber  \\
&=& - g_s \tr {1\over Y-\mathbf R} + g_s\sum_{i=1}^N\sum_{j=0}^\infty {1\over x-j g_s}\,{1\over Y-R_i}\,\left<\left( {1\over j g_s-M}\right)_{i,i}\right>^{(0)}  \nonumber \\
&=& - g_s \tr {1\over Y-\mathbf R} + g_s\sum_{i=1}^N\sum_{j=0}^\infty \sum_{p=0}^\infty  {1\over x-j g_s}\,{1\over Y-R_i}\, T_{j,p;i}\nonumber \\
\eea
$$
T_{j,p;i} = \left<\left( {1\over j g_s-\mathbf{ \tilde R}} \left(\widetilde M\,{1\over j g_s-\mathbf {\tilde R}}\right)^p\right)_{i,i}\right>^{(0)}
$$
where as usual $<.>^{(0)}$ means that we shift $M=\widetilde{\mathbf{R}}+\widetilde M$, and keep only the genus zero fatgraphs in the gaussian expectation value. Notice that the sum over $j$ is absolutely convergent. Moreover, to a given order in $g_s$, the sum over $p$ is finite.

\smallskip

This equation is sufficient to determine $Y(x)$ order by order in $g_s$.
To leading order we find $Y(x)=-x+\ln{(x/t)} + O(g_s)$.
To subleading orders, we recursively have to determine a finite number of coefficients of the $g_s$ expansion of $T_{j,p;i}$. Those coefficients are completely determined by the condition that, to each order in $g_s$, $W_1^{(0)}(x)$ is a rational function of $x$ with poles only at $x=\tilde R_i$, and in particular it must have no pole at $x=j g_s$, or at $Y(x)=R_i$.

\vspace{0.5cm}

\subsection{Asymptotic expansion of the spectral curve}
\label{app:SCform}

To leading order in $g_s$, we have $V'(x)=-x+\ln{(x/t)}$, and:
$$
Y(x)=-x+\ln{(x/t)} + O(g_s)
$$
This leading order spectral curve has a rational uniformization:
$$
\left\{\begin{array}{l}
x(z) = z \\
y(z) = -z+\ln{(z/t)}
\end{array}\right.
$$
Since the higher order $g_s$ corrections to $V'(x)$ are all rational, the corrections to $P_1^{(0)}(x,y)$ are also rational functions of $x$ and $y$.
This implies that all corrections to $Y(x)$ can be written with the uniformizing variable $z$, i.e. to all orders in $g_s$ the spectral curve is of genus $0$.

\medskip

To all orders in $g_s$, the equation Eqn~\ref{loopeqapp} is algebraic of degree $N+1$ in the variable $Y$, this implies that the function $x(z)$ is rational of degree $N+1$. It is easy to see that it has $N$ simple poles at $z_i$ such that $y(z_i)=R_i$, and one simple pole at $\infty$.
Up to a homographical change of variable $z$, we assume that $x(0)=0$ and $x(z)\sim z$ at large $z$, i.e.:
$$
x(z) = z + g_s \sum_{i=1}^N {1\over (x-z_i)\,y_i}+{1\over z_i y_i}
$$
Moreover, one sees directly from Eqn~\ref{loopeqapp}, that the residue of $x\mathrm{d}y$ is $g_s$, i.e. $y_i=y'(z_i)$.
The function $y(z)$ starts to leading order in $g_s$ as $y(z) = -z +\ln{(z/t)}+O(g_s)$, and all the higher $g_s$ corrections are rational functions of $z$.
Let $x=\infty$ and  $X_j=j g_s$, $j\in \mathbb N$ be the singularities of $V'(x)$.
For each $X_j$, let us choose $\hat z_j$ such that $x(\hat z_j)=X_j$ (and we must choose the value of $\hat z_j$ which has a small $t$ power series expansion).
Since $V'(x)$ has a simple pole of residue $g_s$ at $x=X_j$, one sees from Eqn~\ref{loopeqapp}, that  $y\mathrm{d}x$ has a simple pole of residue $g_s$ at $z=\hat z_j$.
Another way of saying this is to write:
$$
y(z) = {1\over 2i\pi}\,\oint_{{\mathcal C}_0} {dz'\over z-z'}\,\, V'(x(z')),
$$
where ${\mathcal C_0}$ is a contour surrounding $\infty$ and all the $\hat z_j$'s.

\medskip

Order by order in $g_s$, using the Stirling expansion of $V'(x)$, this gives the spectral curve ${\mathcal S}(\p;t)$ of Eqn~\ref{eq:SC1}.

\vspace{1.8cm}

\bibliographystyle{amsplain}

\providecommand{\bysame}{\leavevmode\hbox to3em{\hrulefill}\thinspace}
\providecommand{\MR}{\relax\ifhmode\unskip\space\fi MR }
\providecommand{\MRhref}[2]{%
  \href{http://www.ams.org/mathscinet-getitem?mr=#1}{#2}
}
\providecommand{\href}[2]{#2}

\end{document}